\documentclass[10pt, journal,twoside]{IEEEtran} 
\usepackage{stfloats}
\usepackage{graphicx}
\usepackage{subfigure}
\usepackage{epstopdf}
\usepackage{booktabs}
\usepackage{amsmath}
\usepackage{amssymb}
\usepackage{cases}
\usepackage{times} 
\usepackage[font=footnotesize,labelfont=bf]{caption}
\usepackage{caption}
\usepackage{color}
\usepackage{flushend} 
\usepackage{multirow}
\usepackage[T1]{fontenc} 
\usepackage{algorithm, algorithmic}

\makeatletter
  \newcommand\figcaption{\def\@captype{figure}\caption}
  \newcommand\tabcaption{\def\@captype{table}\caption}
\makeatother
\usepackage{setspace}
\usepackage{makecell}
\usepackage{bm}
\usepackage{float}
\usepackage{amsthm}
\usepackage{mathrsfs}
\usepackage{makecell}
\usepackage{pifont}
\usepackage{verbatim}
\usepackage{float} 
\usepackage{amsthm}

\begin{document}
\title{Data Trading Combination Auction Mechanism  based on the exponential mechanism}
\author{Kongyang Chen, Zemin Xu, Bing Mi}
\IEEEtitleabstractindextext{
\begin{abstract}
With the widespread application of machine learning technology in recent years, the demand for training data has increased significantly, leading to the emergence of research areas such as data trading. The work in this field is still in the developmental stage. Different buyers have varying degrees of demand for various types of data, and auctions play a role in such scenarios due to their authenticity and fairness. Recent related work has proposed combination auction mechanisms for different domains. However, such mechanisms have not addressed the privacy concerns of buyers. In this paper, we design a \textit{Data Trading Combination Auction Mechanism  based on the exponential mechanism} (DCAE) to protect buyers' bidding privacy from being leaked. We apply the exponential mechanism to select the final settlement price for the auction and generate a probability distribution based on the relationship between the price and the revenue. In the experimental aspect, we consider the selection of different mechanisms under two scenarios, and the experimental results show that this method can ensure high auction revenue and protect buyers' privacy from being violated.
\end{abstract}
\begin{IEEEkeywords}
Data Trading, Differential Privacy, Auction Mechanism, Revenue
\end{IEEEkeywords}
}
\maketitle
\IEEEdisplaynontitleabstractindextext
\IEEEpeerreviewmaketitle

\section{Introduction}
In recent years, emerging machine learning technologies have achieved success in various learning tasks and are gradually being applied by more and more companies and industries. High-availability machine learning models rely on a large amount of high-quality training data. In the data market, the party holding a large amount of data is often various industry companies. They have a large amount of data with varying quality and limited quantity. Recent research aims to explore how to better utilize this data and deliver it to demanders who require large amounts of data in a reasonable manner.

Research on data trading mainly focuses on data pricing and trading methods. There are many ways to price data. For example, direct trading of the entire dataset \cite{1,2,3,4,5} is not conducive to data owners fully utilizing their data to obtain benefits and loses control over their data. For data buyers, buying the entire dataset is not cost-effective when they only want a portion or specific features of the dataset. Another pricing method is based on queries \cite{6}, where the data set seller charges based on each query specified by the buyer, which partially compensates for the previous two shortcomings. However, this query-based method is often too simple to support complex query situations. In addition, there are trading methods based on model pricing \cite{7,8}.

Currently, many studies on privacy protection have different focuses, and research on buyer privacy protection in particular is still unknown. Our work focuses on studying privacy protection mechanisms that primarily protect the buyer in the context of data transactions. Auctions are rarely used in the field of data transactions, and only a few studies have focused on privacy protection in data transaction auctions. In this paper, we propose a differential privacy-based combinatorial auction mechanism. Choosing a combinatorial auction as the trading method not only allows buyers to obtain the required datasets at reasonable prices but also enables data sellers to provide more datasets and increase their revenue. In an auction scenario, the auction platform is usually assumed to be honest and fair, meaning that the platform will disclose the bids of the bidders during the auction and announce the auction result after the auction ends. However, in a dishonest auction environment, once an honest bidder submits a truthful bid, a selfish seller may participate in the bidding of the same dataset by hiring bidders to increase the final transaction price. This allows dishonest sellers to obtain high profits while harming the interests of honest buyers and not reflecting the true value of the dataset.

To address the issue of privacy leakage, the use of differential privacy \cite{9} has been proposed to protect the bidding privacy of buyers in auction environments. Differential privacy can protect the privacy of published results with strong theoretical guarantees. Currently, there are many studies on differential privacy auctions in different neighborhoods. For example, Frank McSherry \cite{10} and others studied the problem of privacy leakage in auctions with infinite supply. Ruihao Zhu \cite{11} and others proposed a differential private auction mechanism for approximate revenue maximization (DEAR) in the spectrum rental neighborhood. Haiming Jin \cite{12} and others proposed a private incentive mechanism to protect bidders' bidding privacy in the crowdsourcing neighborhood. XuTong Jiang \cite{13} and others proposed a differential privacy-based combined double auction scheme for multi-resource allocation in the edge computing resource trading market. To our knowledge, our work is the first study that applies differential privacy technology to solve the problem of privacy leakage in auction environments in the data trading neighborhood.

The main contributions of our work could be summarized as follows:

\begin{enumerate}
\item To the best of our knowledge, we are the first to apply differential privacy techniques to address buyer bid privacy in the field of data transactions. The system proposed in this paper, DCAE, not only prevents the disclosure of the true bids of the bidders but also ensures both approximate truthfulness and high revenue.
\item Using the exponential mechanism to select the final settlement price vector, with the corresponding income being an exponential proportion, can satisfy the instance constraint while achieving approximate truthfulness.
\item The performance of DCAE was evaluated through extensive experiments, and the experimental results showed that DCAE can generate revenue close to the optimal revenue and protect buyer privacy in different experimental environments.
\end{enumerate}

The rest of this paper is organized as follows. Section~\ref{sec:related} briefly reviews previous related research work. In Section~\ref{sec:architecture}, we introduce the idea and model architecture of this paper. In Section~\ref{sec:design}, we describe the detailed design of DCAE. In Section~\ref{sec:experiment}, we implement the mechanism and evaluate its performance through experiments. Finally, Section~\ref{sec:conclusion} provides a summary and outlook for this paper.

\section{Related Work}\label{sec:related}
In this section, we will discuss related work in the field of data trading on data pricing and data privacy.
\subsection{Data Pricing}\label{sec:data pricing}
There are many studies on pricing methods in the field of data trading. For example, Dawex \cite{1}, Iota \cite{2}, Twitter \cite{3}, Bloomberg \cite{4}, and SafeGraph \cite{5} achieve trading goals by directly selling complete datasets. However, this is not conducive to the data owner's complete control over the data, and this pricing method is not the best choice when data buyers only seek to purchase part of the data. To address this issue, Google Bigquery \cite{6} and others proposed a query-based pricing method, which charges data buyers for a specified query subset of the data to achieve the trading goal, partially mitigating these shortcomings. However, due to the simplicity of the query method and inability to support overly complex queries, this method has limited application scope. Lingjiao Chen \cite{7} and others proposed pricing a group of models based on model quality to maximize seller revenue. Jinfei Liu \cite{8} and others proposed a relatively complete data trading framework based on model pricing.
\subsection{Auction Mechanism}\label{sec:auction mechanism}
Auction is an effective trading method, and there are various auction methods. However, research on auctions in the field of data trading is still limited, and most auction methods are applied in other domains. For example, Wang \cite{14} and others proposed an auction mechanism that can be efficiently computed while maintaining truthfulness. Zaman \cite{15} and others applied a combination auction to the scenario of virtual machine instance supply. In addition, there are many studies on auction mechanisms. Qin \cite{16} and others studied the use of neural networks to learn auction mechanisms.
\subsection{Trading Data Privacy}\label{sec:trading data privacy}
With the rise of data trading, data privacy protection has become an important topic. Research on data trading privacy protection aims to protect personal privacy information in data transactions while ensuring the reliability and effectiveness of data transactions. Zheng \cite{17} and others proposed a blockchain-based data market trading method that uses differential privacy to achieve privacy protection. Zheng \cite{18} and others studied how to empower data owners to control privacy loss and proposed a modular data trading framework with the aim of enabling each data owner to constrain their personal privacy loss in data trading. Niu \cite{19} and others studied how to verify the authenticity of data and privacy protection, using homomorphic encryption and identity-based signatures to preserve both identity and data privacy.

However, the above works mainly focus on privacy protection of data itself during the data transaction process, and the data buyer, as the main role in the transaction process, may suffer privacy leakage in auction scenarios and thus needs privacy protection. In recent years, with the increasing attention to privacy protection issues, research in other domains on this aspect has gradually increased. Chen \cite{20} and others designed a privacy protection cloud auction mechanism using cryptographic techniques, while Cheng \cite{21} and others proposed an efficient privacy-preserving double cloud auction mechanism for bilateral cloud markets, which ensured that information about bids was not leaked. Ni \cite{22} and others conducted a combination auction for virtual machine resources, taking into account the time-limited nature of virtual machine resources, where once the bidder submitted a truthful valuation, attackers may infer some private information of the bidder from the two results. To solve this problem, they used differential privacy to limit the privacy leakage. Currently, differential privacy has been applied to specific auction scenarios, including spectrum auctions \cite{23}, mobile crowdsensing \cite{24}, spectrum sensing \cite{25}, edge computing \cite{26}, and others \cite{27}.

In this paper, unlike virtual machine auctions, we consider malicious attackers in the auction process in the context of data trading. Attackers may use the transparent bids of bidders to maliciously bid and increase the final transaction price, which will harm the interests of the bidders. Therefore, we are the first to use differential privacy techniques to address this issue.

\section{System Architecture}\label{sec:architecture}
Currently, the trading methods of data in the field of data trading are being extensively studied. Previous works have proposed many different trading methods, such as model-based trading, dataset-based trading, and query-based trading, which generally involve buyers bidding, sellers charging fees, and delivering corresponding data. For data sellers, such as some large companies, the datasets they sell often have high data quality and vary in quantity in different scenarios. Such datasets are bound to attract more attention and competition from buyers in the trading market. How to reasonably solve this supply-demand relationship and deliver the dataset to the most competitive buyer in a fair manner is a crucial issue. On the one hand, the most suitable trading method for this type of dataset is combination auction, which allows multiple bidders to simultaneously bid on multiple datasets to achieve the trading goal. On the other hand, in general auction scenarios, sellers act as auctioneers, and because there are selfish sellers in the market, this will expose the buyer's bid, which may lead to malicious sellers hiring bidders to participate in bidding on the same type of dataset to raise the final transaction price, which will not be beneficial to the buyer's interests. Therefore, a fair and reasonable auction mechanism is needed to protect the buyer's bid privacy from being violated while ensuring that the final settlement price is not interfered by the seller.

To achieve this, a fair and dedicated auction platform is needed to conduct the auction, while the exponential mechanism in differential privacy is introduced to protect the buyer's bid privacy. The exponential mechanism is designed for selecting the "best" response, and unlike adding noise to the calculation result, it can achieve privacy protection while obtaining the true result. In addition, unlike traditional auctions, designing a differential privacy-protected auction mechanism in a heterogeneous environment poses two challenges. On the one hand, each bidder can express their preferences for different types of datasets, rather than being limited to one type, so a reasonable pricing and allocation strategy should depend on the types of datasets that bidders bid on. At the same time, the auction mechanism must ensure the data provider's revenue, encourage bidders to submit truthful bids, and not reveal privacy information due to the auction results. Therefore, to address the above issues and challenges, we propose a combination auction mechanism DCAE that considers both honesty and privacy protection in the context of data trading.

Currently, there are various types and methods of data transactions, each with its own strengths, and privacy protection during the transaction process has gradually received attention. The specific privacy protection targets and related technologies also vary. However, research on protecting buyer privacy is still rare. In an auction scenario, given the limited and time-sensitive nature of datasets, buyers with purchasing intentions may offer a high price for a particular dataset. However, dishonest and selfish sellers may hire buyers to bid on the same dataset to artificially increase the final transaction price and profit from it. This is unfair to honest buyers, and transparency of buyer bids allows dishonest sellers to profit from it, making the entire auction process lack honesty and fairness. Therefore, for this scenario, this paper proposes the use of a combinatorial auction form, with the primary target of privacy protection being the buyer, and the differential privacy index mechanism used to protect the privacy of the buyer's bid, thus preventing malicious price inflation by the seller. Furthermore, relevant experiments show that this approach can ensure high revenue while protecting buyer privacy.

\subsection{Overall Architecture}\label{sec:overall architecture}

In this section, we will introduce the specific architecture of the DCAE model, which was proposed in the previous section regarding the auction mechanism. 

The auction model consists of three roles: sellers, auction decision platform, and buyers. We define Sellers as the sellers who sell the dataset, Auction Platform as the auction decision platform, and Buyers as the buyers who purchase the dataset. The auction platform is honest and fair, ensuring the normal operation of the auction process.

In this auction system, the three roles are described as follows:

Sellers: Composed of representatives from companies selling datasets, each seller offers different types of datasets, and each type of dataset may have a limited or unlimited number of copies in different scenarios.

Auction Platform: Responsible for the orderly conduct of the entire auction process, ensuring that the buyer's bid privacy is not leaked and preventing sellers from maliciously driving up prices.

Buyers: Composed of buyers who purchase datasets, each buyer can make a combined bid on datasets sold by multiple sellers, and the specific bid price depends on the buyer's willingness to purchase the dataset.

The architecture diagram of the auction process is shown in Figure 1 below:
\begin{figure}[!htp]
\centering
\includegraphics[width=0.5\textwidth]{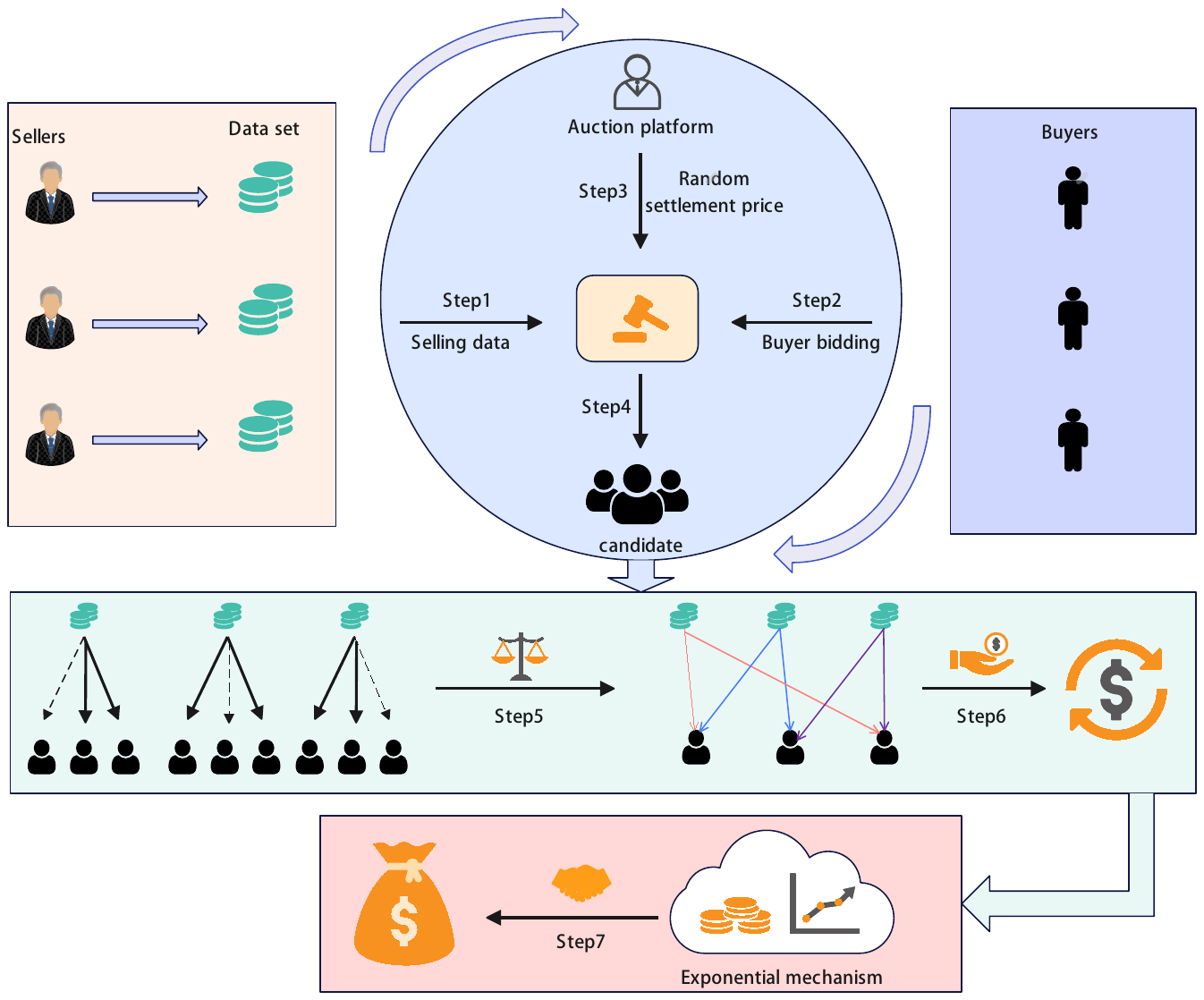} 
\caption{DCAE Transaction Process.}
\label{fig:figure_1}
\end{figure}

Step 1: Sellers publish their owned datasets to the auction platform. Each seller has different types of datasets, and each dataset has a limited number of copies.

Step 2: Buyers make bids on datasets they are interested in based on the datasets published by sellers on the auction platform. The bid includes the bid unit price and bid quantity for each dataset.

Step 3: The auction platform generates a random settlement price vector in each round of the auction process to ensure the honesty and fairness of the auction.

Step 4: Based on the settlement price vector generated and the bids submitted by buyers, the auction platform selects a candidate set of bidders that meet the criteria according to a specific pricing strategy.

Step 5: After the candidate set is selected, the auction platform randomly assigns the bidders in the candidate set and chooses the final winner under the constraint of the bid quantity.

Step 6: The auction platform determines the final revenue of the auction round based on the winner and repeats Steps 1-6.

Step 7: Based on the results of multiple rounds of auctions, an exponential mechanism is used to generate a probability distribution about the settlement price and final revenue. The exponential mechanism selects the optimal revenue as the final auction result while protecting the privacy of buyer bids.

In the above model, there are issues such as how to develop appropriate pricing strategies for selecting candidates and how to allocate datasets to winners to ensure honesty while maintaining differential privacy protection on the pricing process. Next, we will detail the modeling process of the above flowchart and the solutions to the existing problems.

\section{Detailed Design}\label{sec:design}
\subsection{Parameter And Symbol Setting}
Regarding the DCAE model proposed in the previous chapter, we assume that \(T\) sellers have published \(m\) types of datasets to the auction platform, and the set of these \(m\) datasets is denoted as \(M = \left\{ D_{1},D_{2},D_{3}\ldots D_{m} \right\}\).The number of each type of dataset is different, and \(K = \left\{ K_{1},K_{2},K_{3}\ldots K_{m} \right\}\) represents the set of the number of copies of each type of dataset, where the number of the \(i-th\) type of dataset is denoted as \(K_i\). We assume that \(n\) buyers bid on the \(m\) types of datasets through the auction platform, and the set of buyers is denoted as \(N = \left\{ 1,2,\ldots,j,\ldots,n \right\}\). The bid of the \(j-th\) buyer for the dataset is denoted as \(B = < k_{j},b_{j} >\), where \(k_{j} = \left\{ k_{j}^{1},k_{j}^{2},\ldots,k_{j}^{i},\ldots,k_{j}^{m} \right\}\) represents the set of the number of copies of the \(m\) types of datasets requested by the \(j-th\) buyer, and \(K_{j}^{i} \in \left\lbrack 0,q_{max} \right\rbrack\). If a buyer requests a large amount of datasets, there is no need to participate in the auction and they can directly purchase from the corresponding seller, so \(q_{max} < < K_{i}. b_{j} = \left\{ b_{j}^{1},b_{j}^{2},\ldots,b_{j}^{i},\ldots,b_{j}^{m} \right\}\) represents the set of the unit prices submitted by the \(j-th\) buyer for the \(m\) types of datasets. We assume that every buyer is rational, that is, if the \(j-th\) buyer is not interested in the \(i-th\) dataset, \(k_{j}^{i} = 0,b_{j}^{i}\) = 0. Therefore, the total bidding price of all datasets requested by the \(j-th\) buyer is \({\overset{-}{B}}_{j} = {\sum\limits_{i = 1}^{m}k_{j}^{i}}b_{j}^{i}\), and the set of bids for all buyers participating in the auction is denoted as \(B = \left\{ B_{1},B_{2},\ldots,B_{n} \right\}\). After the auction, the final results include an allocation set \(x = \left\{ x_{1},x_{2},\ldots,x_{j},\ldots,x_{n} \right\}\) and a payment set, where \(x_{j}\) represents whether the \(j-th\) buyer wins and \(P_{j}\) represents the final settlement price of the \(j-th\) buyer. We assume that every buyer is rational, meaning they will only bid on datasets they are interested in and will not bid on datasets they are not interested in. Therefore, the corresponding settlement price is 0, that is, \(P_{j} = 0\). The auction platform needs to maximize its final revenue target, which is the total amount paid by all buyers, that is, \(REV = {\sum\limits_{j = 1}^{m}{P_{j}x_{j}}}\).

\subsection{Combination Auction Mechanism}
The combinatorial auction mechanism is a type of auction format that allows buyers to simultaneously bid on combinations of multiple goods or services at a fixed price. The distinctive feature of combinatorial auction mechanism is that in one auction, buyers can select combinations of goods or services that interest them and bid to win those combinations. Sellers can choose some combinations of goods or services to sell and set a reserve price. In the end, the buyer who bids the highest among all the bidders will win the selected combination, and the seller will receive the total price of that combination.
The advantages of the combinatorial auction mechanism include meeting the diverse needs of buyers, increasing the sales volume and profits of sellers, and reducing the complexity and time cost of auctions.
In auctions, the assurance of truthfulness is related to dominant strategy [28]. Truthfulness means that bidders will provide their true bids, but in some cases, it may not meet the exact truthfulness due to too many restrictions. Therefore, an approximate version of truthfulness, namely, approximate truthfulness [29], should be considered.
\subsection{Differential Privacy}
Let \(B\) denote the set of bids from each user for the desired data set. Then, the combinatorial data set auction mechanism \(M\) satisfies differential privacy if and only if for any two bid sets \(B\) and \(B'\) that differ in only one bid, and for any subset \(S \subseteq Range(M)\), the following holds:
\begin{equation}
P~r\left\lbrack M(B)~ \in ~S \right\rbrack~ \leq ~exp(\epsilon)~ \times ~P~r\left\lbrack M(B')~ \in ~S \right\rbrack
\end{equation}
The privacy budget \(\epsilon\) is a parameter in the definition of differential privacy, which is used to adjust the degree of privacy protection that differential privacy can provide. A smaller \(\epsilon\) means a higher degree of privacy protection that can be provided, while a larger \(\epsilon\) means a lower degree of privacy protection that can be provided.
\subsection{Exponential Mechanism}
The exponential mechanism [10] is a privacy-preserving technique in differential privacy [9],[30]. The basic mechanisms of differential privacy include the Laplace mechanism and the Gaussian mechanism, which are aimed at numerical results by adding noise to the numerical results, with specific noise distributions varying according to the selected mechanism. The exponential mechanism is suitable for cases where specific results are required while ensuring that the response process satisfies differential privacy. This mechanism selects the best result from all the candidate results while ensuring differential privacy. To implement this mechanism, a set of candidate results and a scoring function need to be defined. The scoring function is used to score each result in the set of candidate results and output the corresponding score, and the highest score corresponds to the best result. The exponential mechanism achieves differential privacy by returning a result with a score that approximates the best score. Therefore, the score corresponding to the result returned by the exponential mechanism may not be the highest score in the set of candidate results. Therefore, the process by which the exponential mechanism satisfies \(\epsilon\)-differential privacy is as follows:
\begin{enumerate}
\item Select a set of candidate results \(R\).
\item Specify a scoring function \(\left. u:D~ \times ~R~\rightarrow~R \right.\) with a global sensitivity of \(\mathrm{\Delta}u\).
\item The exponential mechanism outputs \(r \in R\), and the output probability of each response is as follows: \(exp\left( \frac{\epsilon u(x,r)}{2\mathrm{\Delta}u} \right)\).
\end{enumerate}
\subsection{Pricing Strategy}
In the above-designed auction scenario, there are often multiple Buyers participating. To ensure the honesty and high revenue of the entire auction process, a reasonable pricing strategy needs to be designed. Traditional pricing strategies include total price pricing and average unit price pricing. The first pricing method selects the winner among Buyers whose total bid is higher than the total price. This pricing is beneficial to Buyers who request more data sets but ignores Buyers who request fewer data sets but have a higher unit bid, resulting in lower auction revenue. The second pricing method selects the winner among Buyers whose average unit bid is higher than the average unit price. This pricing is instead beneficial to Buyers with higher unit bids, and may result in data sets with lower bids not being allocated, leading to lower revenue. In addition, since the same total price or average unit price may correspond to different combinations of unit bids and number of data sets, implementing these two pricing strategies with honesty is challenging. To address this issue, this system adopts a two-tier pricing approach, as follows:
\begin{enumerate}
\item Tier 1 pricing: based on the unit price of each data set, a unit price vector is constructed, which provides a universal pricing form and ensures fairness in the auction process.
\item Tier 2 pricing: the total price related to each Buyer is calculated based on the bids provided by the Buyer, and compared with the total price calculated based on the unit pricing vector. The result of this pricing is that the price paid by the Buyer is proportional to the number of data sets they request. Figure 2 provides an example comparison of the three pricing strategies, from which it can be concluded that the two-tier pricing provides a reasonable pricing strategy.
\end{enumerate}

\begin{figure}[!htp]
\centering
\includegraphics[width=0.5\textwidth]{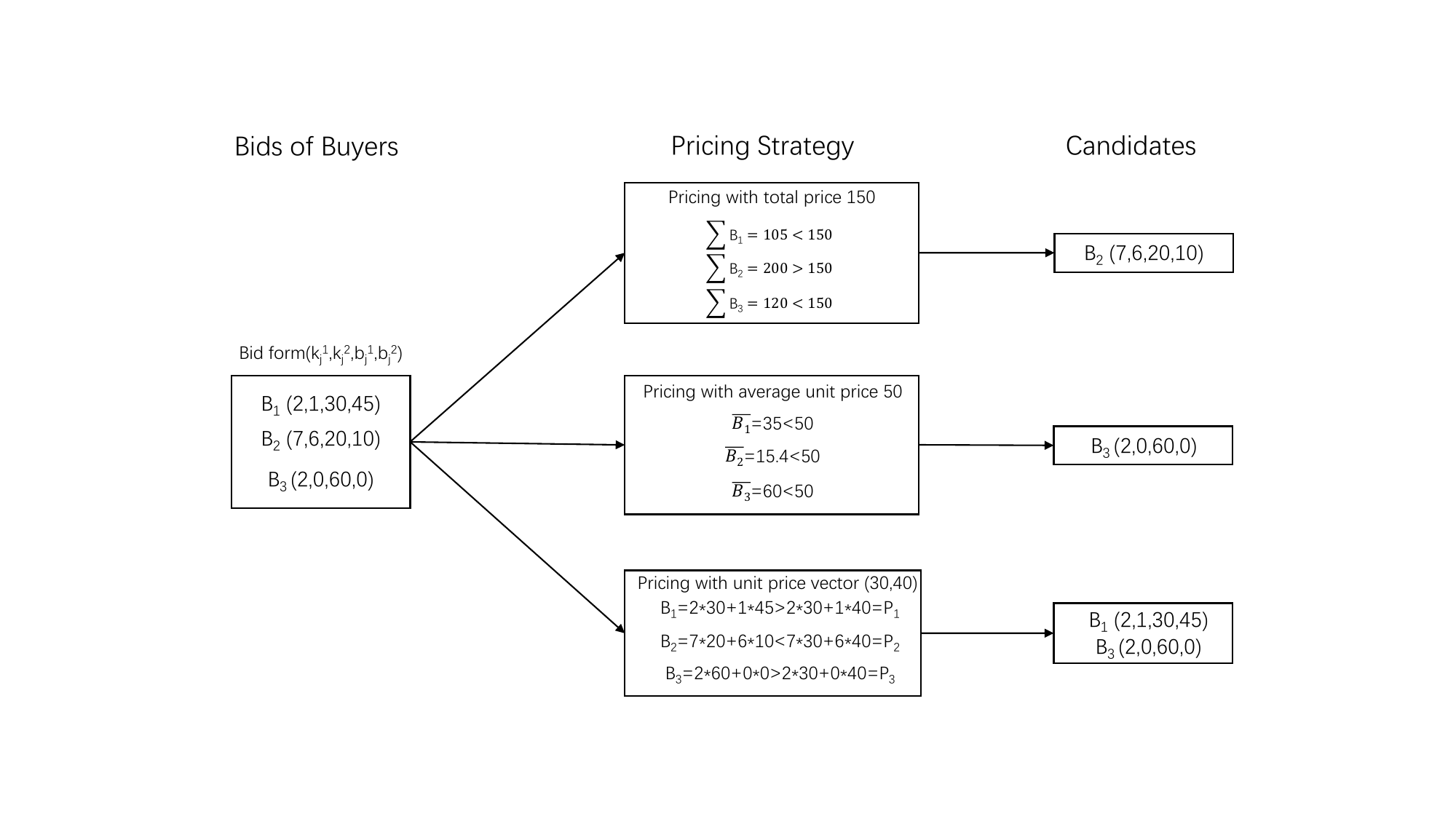} 
\caption{Figure 2 Example comparison of three pricing strategies.}
\label{fig:figure_2}
\end{figure}

\subsection{Distribution Of Winners}
The final question is how to allocate the data sets to the ultimate winners based on the candidate set, while ensuring honesty and pricing with differential privacy protection. Since the number of data sets is limited, not every candidate can win. To achieve honesty, the auction platform should randomly assign the data sets to the candidates, ensuring that the allocation is independent of their bids. This leads to probabilistic revenue, meaning that any settlement price vector may correspond to multiple revenue values. However, using the exponential mechanism to select a settlement price vector requires that the corresponding revenue is deterministic. To address this issue, the auction platform needs to generate a random allocation sequence before the auction and use the exponential mechanism under this random allocation sequence to select the settlement price vector and the final revenue. Therefore, for the auction platform, the final allocation sequence is known, and the revenue is deterministic, while for the Buyers, the allocation sequence is random, meaning that the sequence of winners is random.
\subsection{Detailed Procedure}
Based on the design in the above sections, the final settlement price vector ensures fairness and introduces the exponential mechanism to ensure that the Buyers' bids are not leaked. The final settlement price vector is represented by \(\mathbf{p} = \left\{ p_{1},p_{2},p_{3},\ldots,p_{m} \right\}\), where \(p_i\) represents the unit settlement price of the \(i-th\) dataset. DCAE generates multiple combination lists about \(\mathbf{p}\) and \(REV\) through multiple rounds of auction iteration, and to protect the privacy of Buyers' bids, the exponential mechanism is applied to select the final unit settlement price vector. The specific process is as follows:

In each round of the auction process, the auction platform randomly generates a settlement price vector \(\mathbf{p}\) and selects eligible candidates according to the two-tier pricing strategy. That is, given the settlement price vector \(\mathbf{p}\), the payment price of the \(j-th\) Buyer is:
\begin{equation}
P_{j} = {\sum\limits_{i = 1}^{m}{k_{j}^{i}p_{i}}}
\end{equation}

For the payment price, if the total bid of the \(j-th\) Buyer is not less than the total payment price corresponding to \(\mathbf{p}\), then the Buyer is selected as a candidate.
\begin{equation}
{\overset{-}{B}}_{j} \geq P_{j}
\end{equation}
The final candidate set is denoted as \(W_c\).

Final allocation of data sets.
To ensure that the candidate's winning is completely independent of their bids, the auction platform needs to randomly sort the candidate set, which is important for achieving honesty. To ensure the fairness of the auction platform, in each round of the auction, the auction platform randomly generates an allocation sequence \(r\) for the candidate set to determine the allocation order, thus ensuring the fairness of the allocation. Under this condition, the first Buyer who satisfies the data set quantity constraint, i.e., the quantity of each data set requested by the Buyer does not exceed the quantity of the provided data set, is selected as the winner. At the same time, the remaining available data set quantity is updated, and for the remaining candidates, the above process is repeated until all candidates are selected or the available data set quantity is zero, resulting in the winner set \(W\). Based on the winner set, the total revenue of the auction can be calculated as follows:
\begin{equation}
REV\left( {B,K,r,\mathbf{p}} \right) = {\sum\limits_{j \in W}{\sum\limits_{i = 1}^{m}{p_{i}k_{j}^{i}}}}
\end{equation}

Exponential mechanism for price selection.
In this auction model, we consider the set of candidate result sets \(R\) as the set of all possible settlement price vector sets, denoted as: \(\mathbf{p} \in \Pi\), and the scoring function u corresponds to the auction revenue associated with each settlement price vector, which is represented by \(REV\left( {B,K,r,\mathbf{p}} \right)\) as mentioned earlier in the text.
Given the settlement price vector \(\mathbf{p}\) and the random allocation order r, repeat steps 
xxxxxxxxxxxxxxxxxxxxxxxxxxxx
to obtain all possible price vectors and their corresponding revenues. To achieve privacy protection and better revenue, use the exponential mechanism to select the settlement price vector, and calculate the probability distribution of the settlement price vector according to the following formula.
\begin{equation}
{\Pr\left( {M\left( {B,K,r} \right) = \mathbf{p}} \right)} = \frac{exp\left( \frac{\varepsilon REV\left( {B,K,r,\mathbf{p}} \right)}{2\mathrm{\Delta}} \right)}{\sum\limits_{\mathbf{p}^{'} \in \Pi}{exp\left( \frac{\varepsilon REV\left( {B,K,r,\mathbf{p}^{\mathbf{'}}} \right)}{2\mathrm{\Delta}} \right)}}
\end{equation}

\section{Experiment Results}\label{sec:experiment}
This chapter will demonstrate the feasibility and performance of DCAE through experiments.
\subsection{Non-competitive scenario}
In a non-competitive scenario, there are a total of m types of datasets distributed by sellers to the auction platform, and there are a limited number of each type of dataset. In this scenario, most of the dataset quantities can meet the buyers' requests. Assuming \(m = 6,n = 100,k_{i} \in \lbrack 0,100\rbrack,b_{i} \in \lbrack 1,20\rbrack,p \in \lbrack 1,100\rbrack\). Through 100 experiments, each experiment will conduct 1000 auction iterations to obtain multi-round results, and the best result will be selected through an exponential mechanism in each experiment.

Figure 3 compares the revenue of DCAE and Random in 100 experiments.
\begin{figure}[!htp]
\centering
\includegraphics[width=0.5\textwidth]{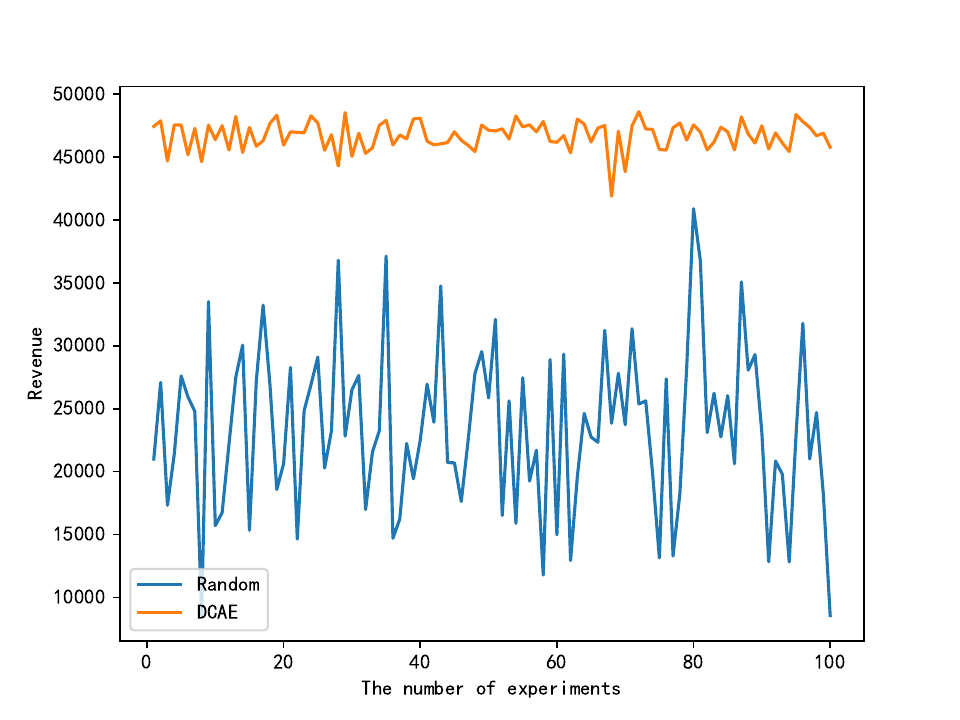} 
\caption{Revenue comparison in non-competitive scenario.}
\label{fig:exp_1}
\end{figure}

As shown in Figure 3, the auction revenue obtained by Random fluctuates greatly in each experiment, with a large variance, while the revenue obtained by DCAE is relatively stable with a smaller variance, and its mean value is maintained at a higher level than Random. This means that DCAE can provide better revenue guarantee compared to Random.
Figure 4 controls the monotonically increasing change of m and compares the changes in the mean revenue and mean satisfaction of the two mechanisms in every 100 experiments. Satisfaction is defined as the proportion of winners to all buyers, i.e., \(\frac{|W|}{n}\).

\begin{figure}[h]
\centering
\subfigure[]{
\label{exp2.sub.1}
\includegraphics[width=0.45\textwidth]{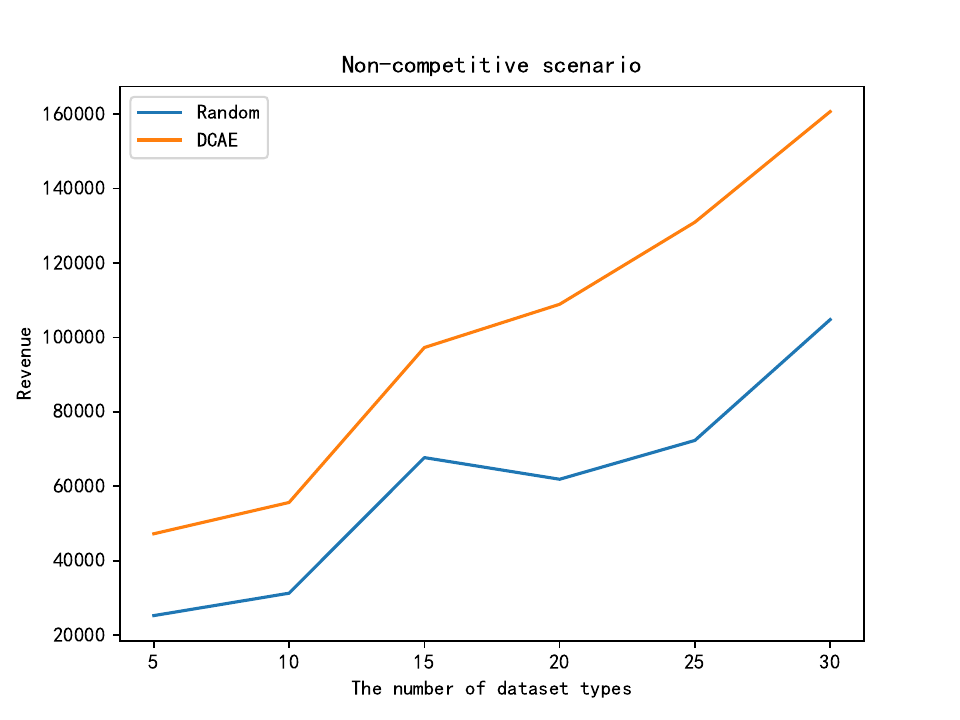}}
\subfigure[]{
\label{exp2.sub.2}
\includegraphics[width=0.45\textwidth]{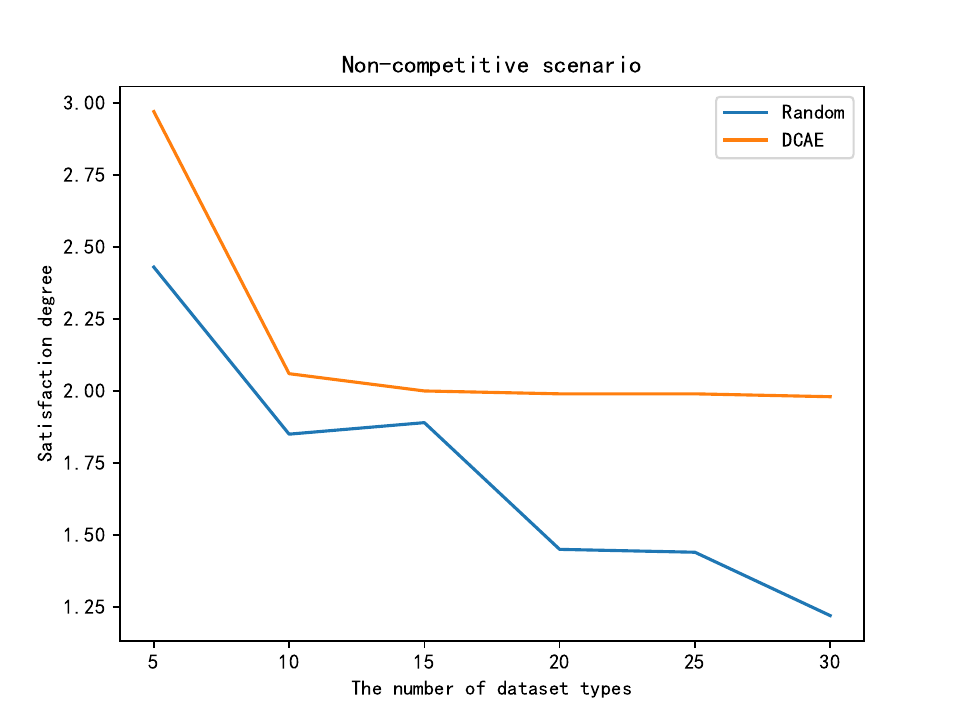}}
\caption{Mean revenue change (a) and mean satisfaction change (b) in Figure 4}
\label{fig:exp_2}
\end{figure}

Figure 4 shows the changes in the mean revenue and satisfaction over 100 experiments as the number of dataset types varies in the range of \(m \in [5, 30]\). From Figure (a), it can be seen that as the number of types increases, DCAE shows a monotonically increasing trend, indicating that DCAE can also guarantee higher revenue when there are more dataset types, which helps attract more sellers to publish data on the auction platform. On the other hand, although Random maintains an increasing trend, its growth rate is smaller compared to DCAE, meaning that the final revenue is lower, and the auction platform cannot obtain optimal revenue. From Figure (b), as the number of types monotonically increases, DCAE's satisfaction mean always stays above Random's. As the probability of datasets that meet buyers' requirements increases with the increase of dataset types, buyers will offer higher prices for datasets that meet their needs. The increase in buyers offering high prices will inevitably affect the number of final candidates and may eliminate more buyers, leading to a decrease in the number of buyers who ultimately win and a decrease in satisfaction. Therefore, as the number of types increases, DCAE's satisfaction curve shows a downward trend at the beginning and then gradually converges. As mentioned earlier, \(b_{i} \ll p_{i}\) (the buyer's bid is much less than the settlement price), so more buyers are needed to support a better revenue. Therefore, the exponential mechanism selects the final settlement vector according to the probability distribution of the optimal revenue each time, ensuring high revenue while maintaining the stability of the number of winners, and preventing the satisfaction curve from decreasing. Overall, DCAE can make more buyers successfully bid while increasing the number of dataset types, while Random only leads to fewer buyers winning as the number of types increases, resulting in a continuous decline in the curve.

Figure 5 shows the changes in mean revenue and mean satisfaction obtained in every 100 experiments as \(K \in [200,800]\).

\begin{figure}[h]
\centering
\subfigure[]{
\label{exp3.sub.1}
\includegraphics[width=0.45\textwidth]{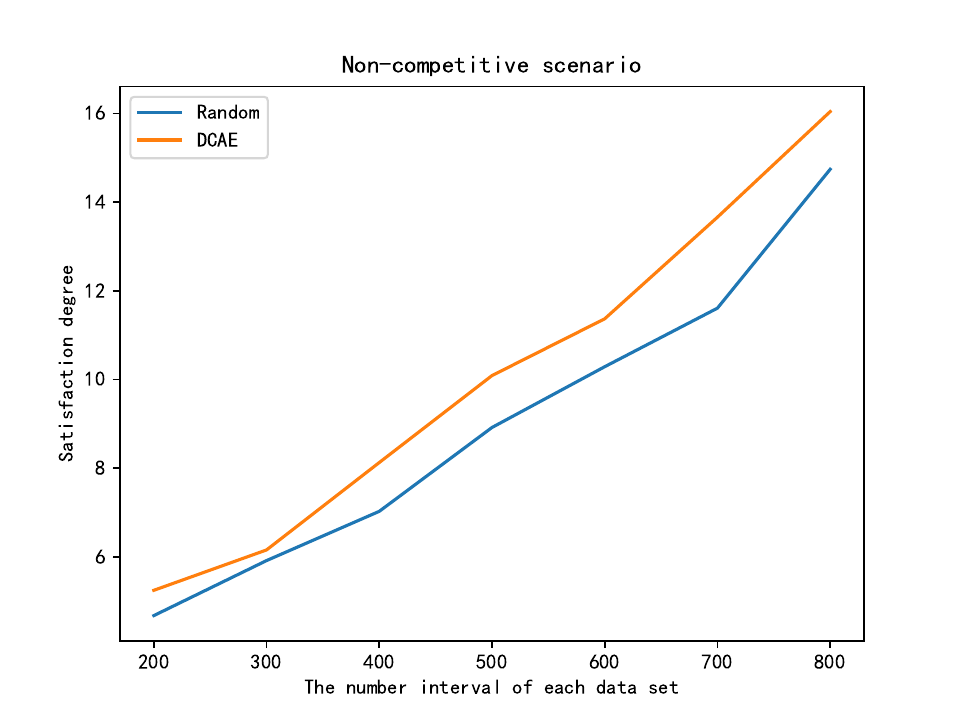}}
\subfigure[]{
\label{3xp3.sub.2}
\includegraphics[width=0.45\textwidth]{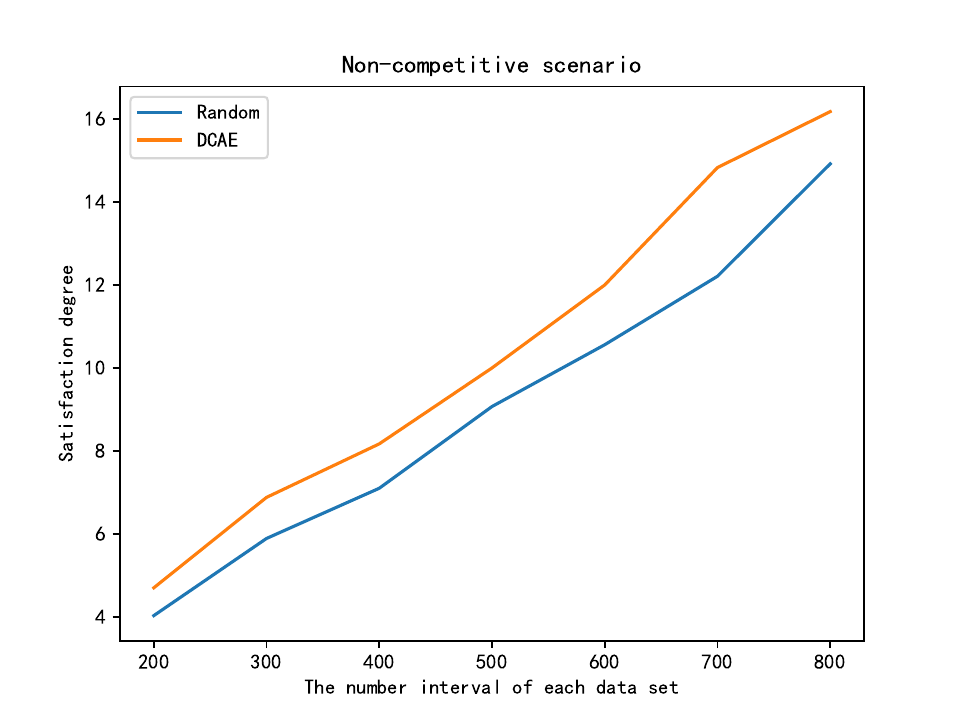}}
\caption{Mean income change (a) and mean satisfaction change (b)}
\label{fig:exp_3}
\end{figure}

From Figure A, it can be observed that as the number of datasets increases, DCAE exhibits an overall increasing trend, while the curve for Random is relatively flat, and its peak revenue is around 200,000, whereas DCAE's peak revenue is close to 300,000, indicating a significant difference between the two. According to Figure B, as the number of each type of dataset increases, more buyers' needs can be met, resulting in more winners, and DCAE always maintains an overall increasing trend, with a higher rate of increase than Random. This means that as sellers publish more data, DCAE can ensure that the auction platform achieves high revenue and satisfaction levels.

To study the level of privacy protection provided by the DCAE mechanism and the revenue obtained under different privacy levels, we compared the performance of DCAE, Random, and the best revenue obtained in each experiment. Figure 6 shows the changes in revenue and mean satisfaction for every 100 experiments as the privacy budget \(\epsilon\) varies within the range of \([0.2, 1]\), where a higher privacy budget indicates a higher level of privacy protection.

\begin{figure}[h]
\centering
\subfigure[a]{
\label{exp4.sub.1}
\includegraphics[width=0.45\textwidth]{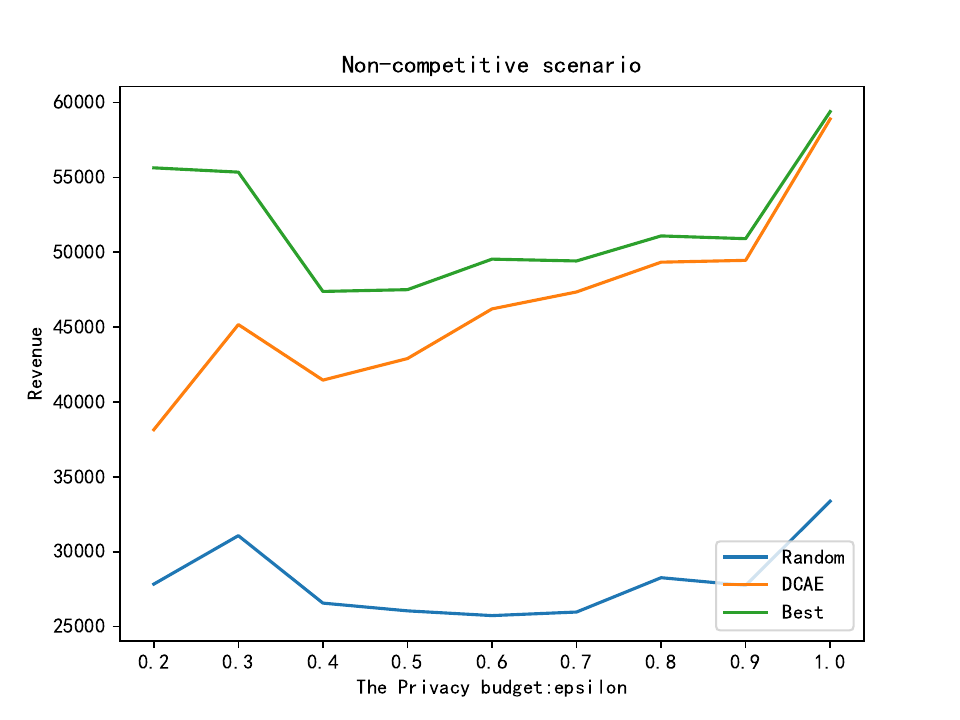}}
\subfigure[b]{
\label{exp4.sub.2}
\includegraphics[width=0.45\textwidth]{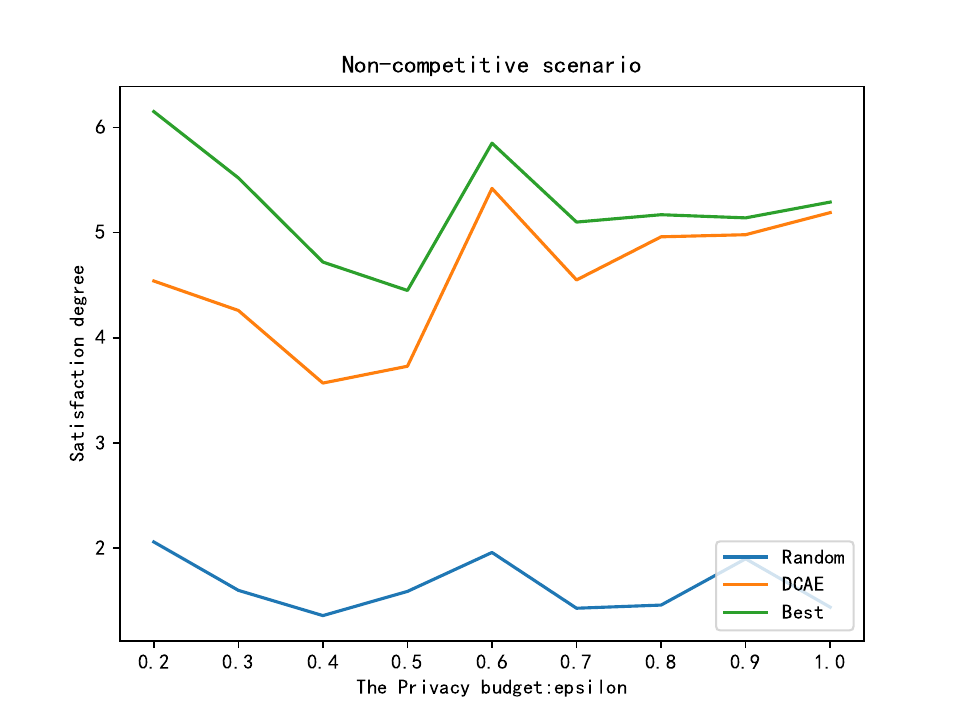}}
\caption{Changes in Mean Revenue (a) and Mean Satisfaction (b)}
\label{fig:exp_4}
\end{figure}

From Figure 6, it can be observed that for the revenue performance shown in Figure A, as the privacy budget increases, the amount of noise in the DCAE output results decreases. Therefore, the accuracy of the output results also improves, indicating that DCAE approaches Best. This trend can be seen from Figure A, where as epsilon increases, the level of privacy protection increases, resulting in a decrease in the degree of closeness between DCAE and Best. The curve for Random, on the other hand, is not affected by epsilon and cannot guarantee higher revenue results. Similarly, from an analysis of Figure A, as the level of privacy protection increases, DCAE approaches Best, making the number of winners for each closer and resulting in a closer level of satisfaction. This trend can be seen from the curve in Figure B, while Random is not affected by the level of privacy budget and cannot result in more winners, resulting in an overall lower satisfaction curve.

Figure 7 shows a comparison between the highest and lowest revenue obtained in each experiment and the optimal revenue selected by DCAE.

\begin{figure}[h]
\centering
\subfigure[]{
\label{exp5.sub.1}
\includegraphics[width=0.45\textwidth]{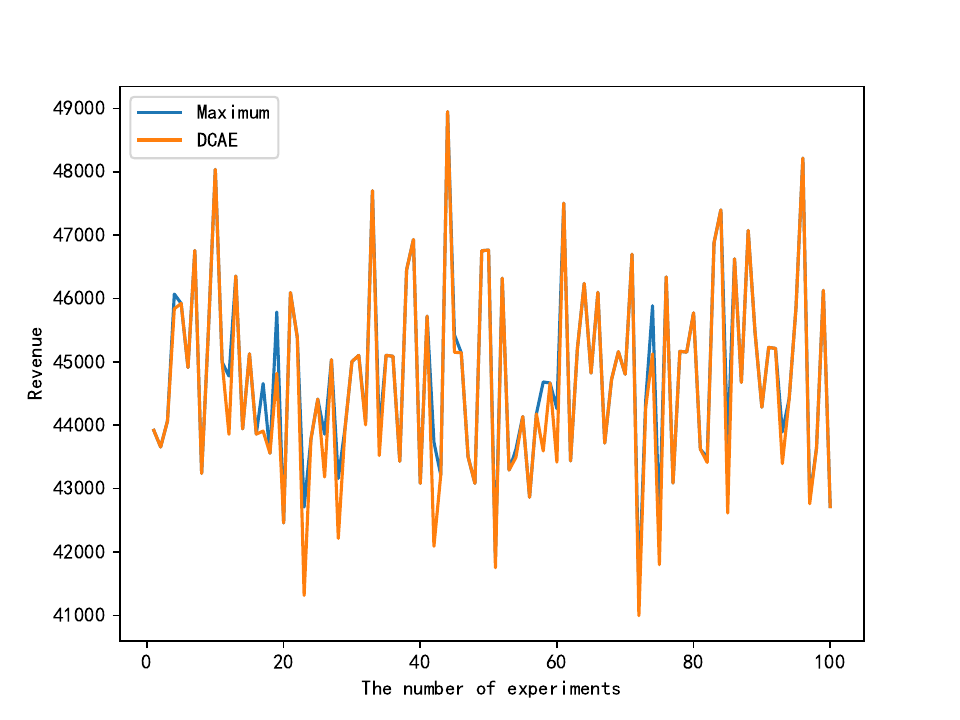}}
\subfigure[]{
\label{exp5.sub.2}
\includegraphics[width=0.45\textwidth]{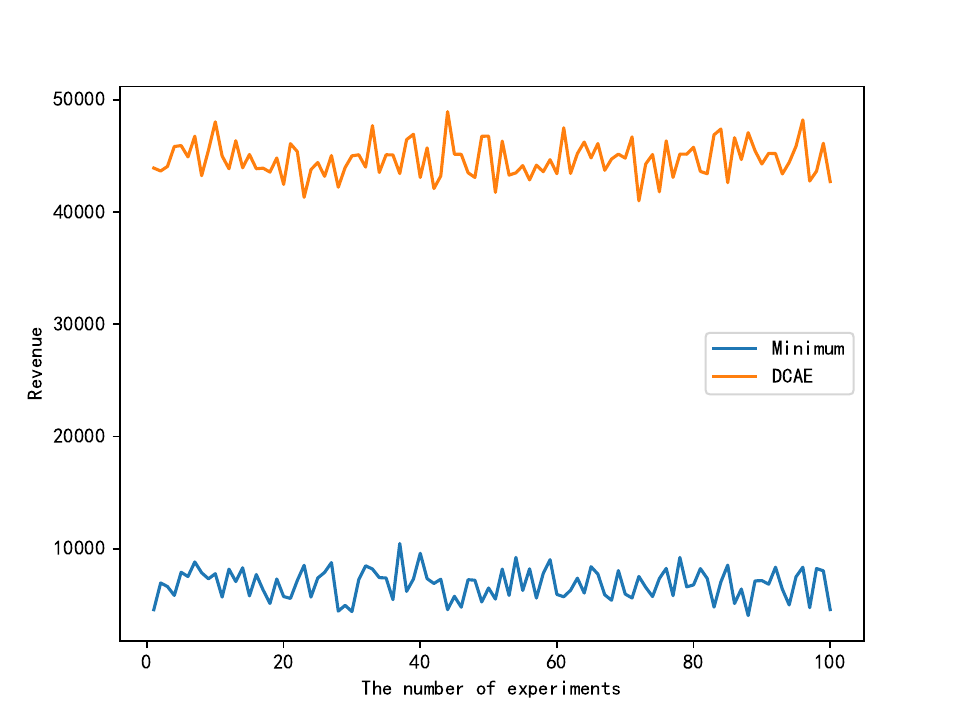}}
\caption{Difference from Optimal Revenue (a) and Difference from Lowest Revenue (b)}
\label{fig:exp_5}
\end{figure}

From Figure 7, it can be observed that DCAE can ensure results that are close to the maximum revenue, while maintaining a significant difference from the lowest revenue. This means that DCAE can choose the optimal revenue close to the maximum revenue while ensuring the level of privacy protection in each experiment, without selecting lower revenue.
\subsection{Competitive scenario}
In a competitive scenario, there are a total of m types of datasets distributed by Sellers to the auction platform, and there is only one copy of each type of dataset. In this scenario, the majority of dataset types cannot meet the requests of Buyers. Figure 8 compares the revenue performance of two mechanisms under this scenario, with other parameters set the same as in the non-competitive scenario.

\begin{figure}[h]
\centering
\subfigure[]{
\label{exp6.sub.1}
\includegraphics[width=0.45\textwidth]{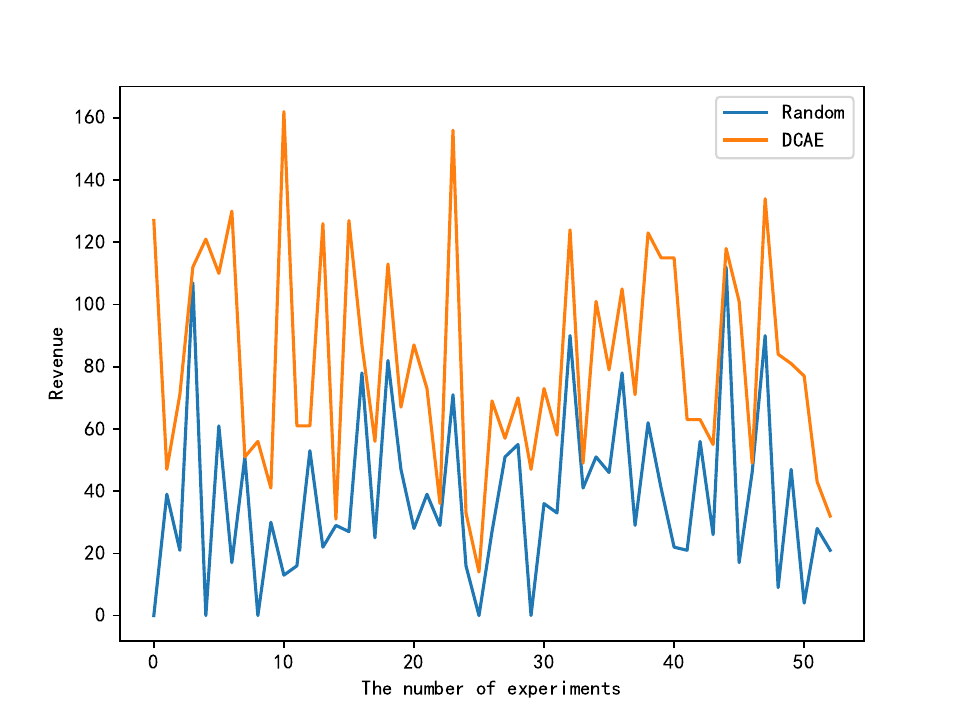}}
\subfigure[]{
\label{exp6.sub.2}
\includegraphics[width=0.45\textwidth]{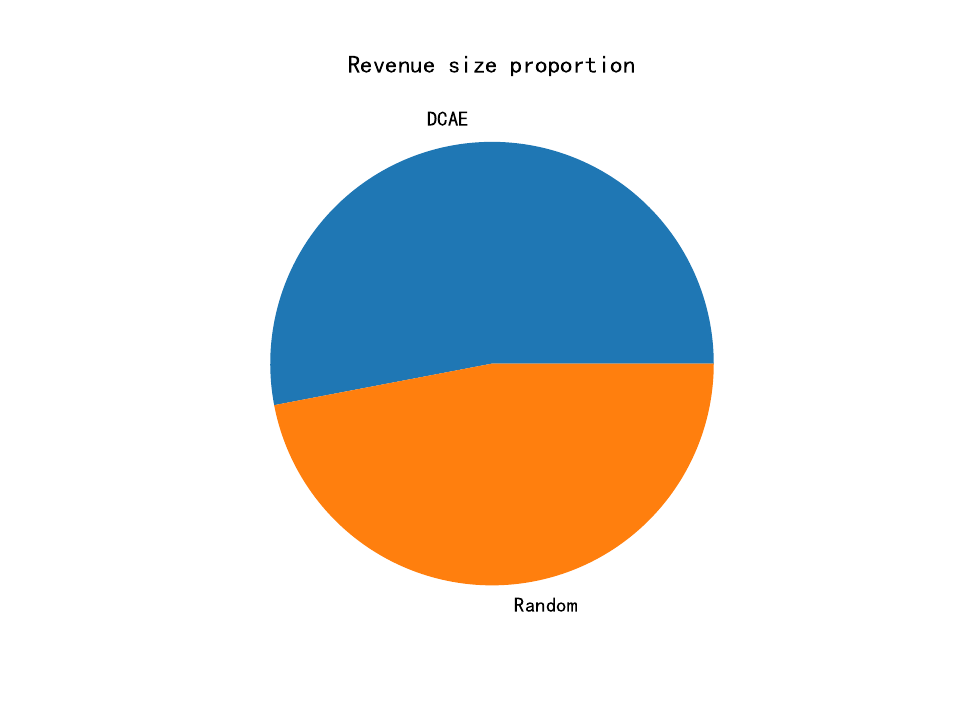}}
\caption{Revenue comparison in the competitive scenario}
\label{fig:exp_6}
\end{figure}

From Figure 8, it can be observed that in the competitive scenario, DCAE still outperforms the Random mechanism in terms of revenue obtained in each experiment. This provides DCAE with a good lower limit guarantee, indicating that even when there is only one copy of each dataset type, DCAE can still ensure high revenue in the final outcome.

Figure 9 compares the changes in mean revenue and mean satisfaction for both mechanisms over every 100 experiments, by controlling a monotonically increasing value of m within the range of [5,30].

\begin{figure}[h]
\centering
\subfigure[]{
\label{exp7.sub.1}
\includegraphics[width=0.45\textwidth]{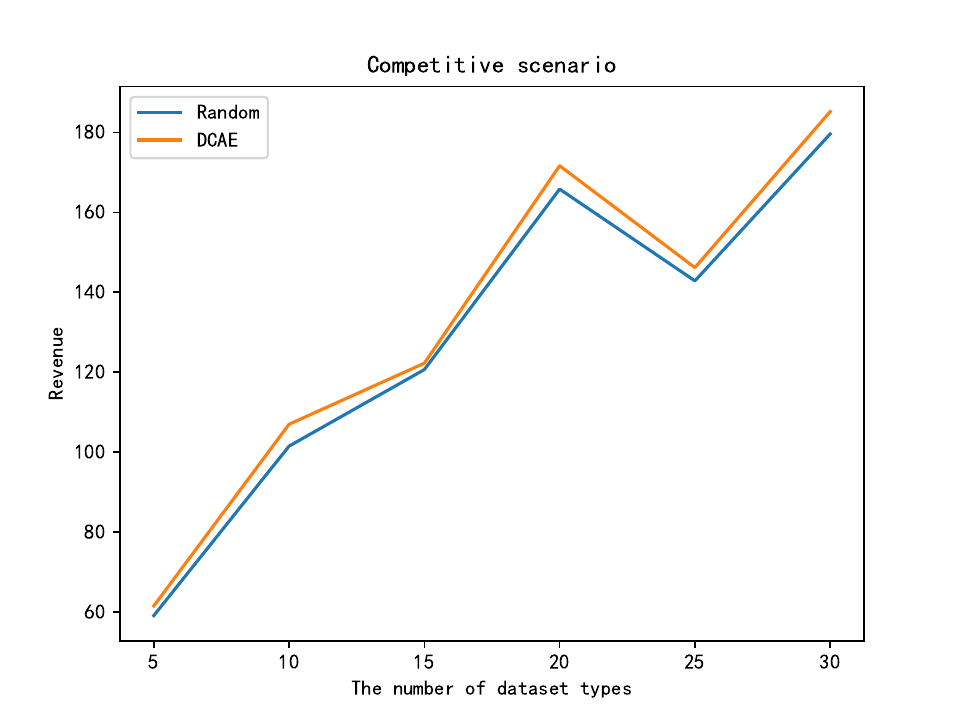}}
\subfigure[]{
\label{exp7.sub.2}
\includegraphics[width=0.45\textwidth]{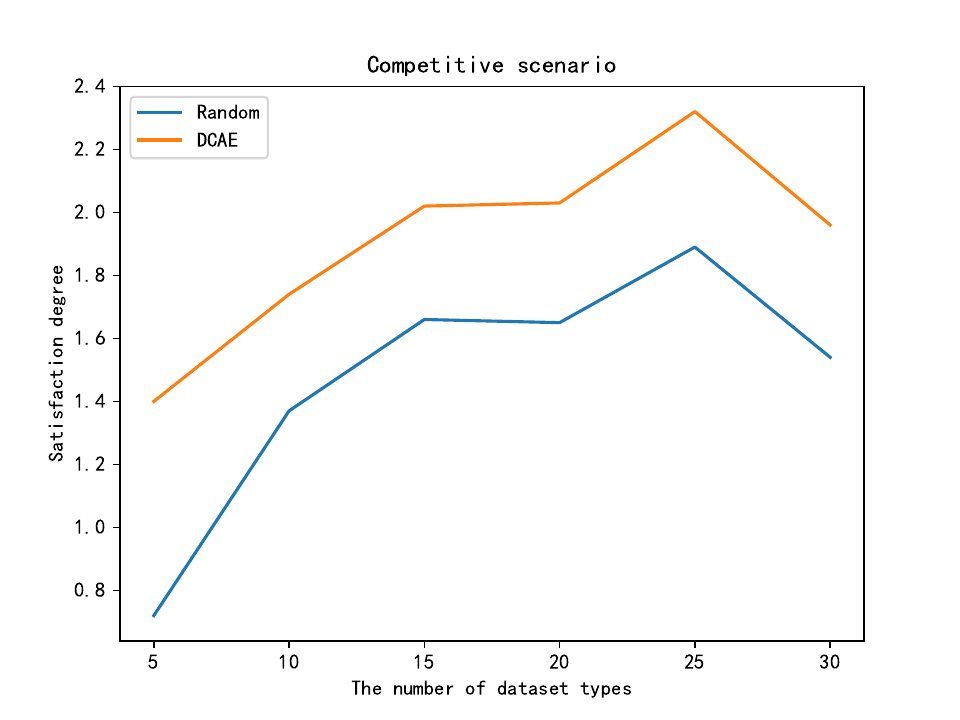}}
\caption{Changes in Mean Revenue (a) and Mean Satisfaction (b)}
\label{fig:exp_7}
\end{figure}

From Figure 9, it can be seen that in a competitive scenario, as the number of dataset types increases, the DCAE and Random mechanisms have similar effects on revenue, and the mean revenue is lower compared to the competitive scenario. Under the limitation of dataset quantity, although both curves show a fluctuating upward trend, DPCA can still maintain higher revenue than Random. In terms of satisfaction, both mechanisms show a trend of increasing, decreasing, and then increasing again, indicating that as the number of dataset types increases, due to the limitation of only one copy of each dataset, most buyers' requests cannot be satisfied, resulting in a decreasing trend of satisfaction.

Figure 10 shows the changes in mean revenue and mean satisfaction for every 100 experiments as the privacy budget epsilon varies within the range of [0.2,1], where a higher privacy budget indicates a higher level of privacy protection.

\begin{figure}[h]
\centering
\subfigure[]{
\label{exp8.sub.1}
\includegraphics[width=0.45\textwidth]{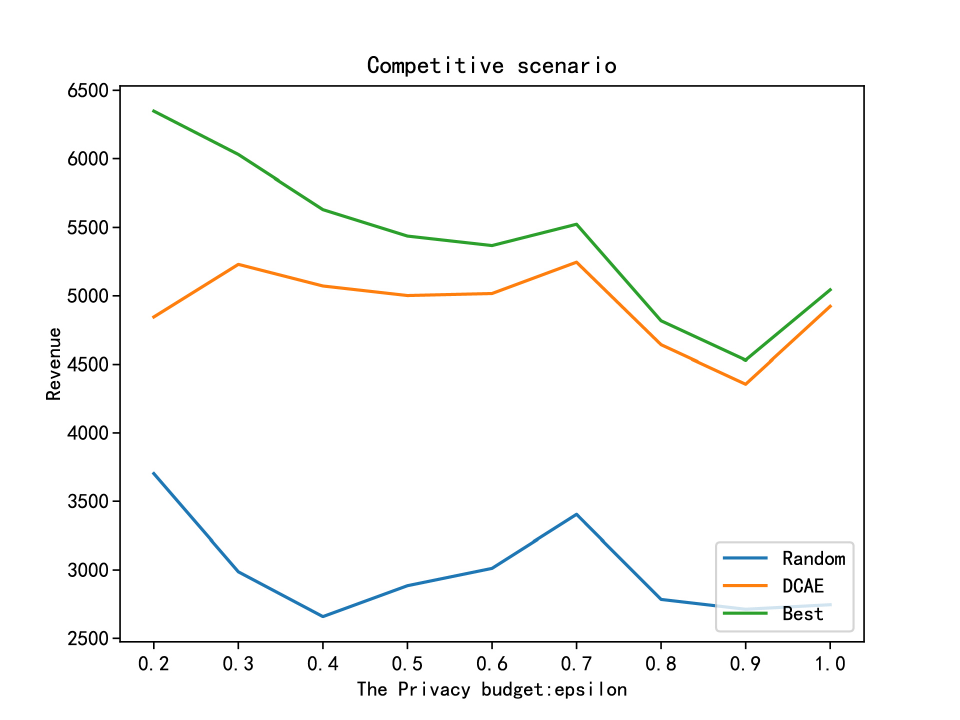}}
\subfigure[]{
\label{exp8.sub.2}
\includegraphics[width=0.45\textwidth]{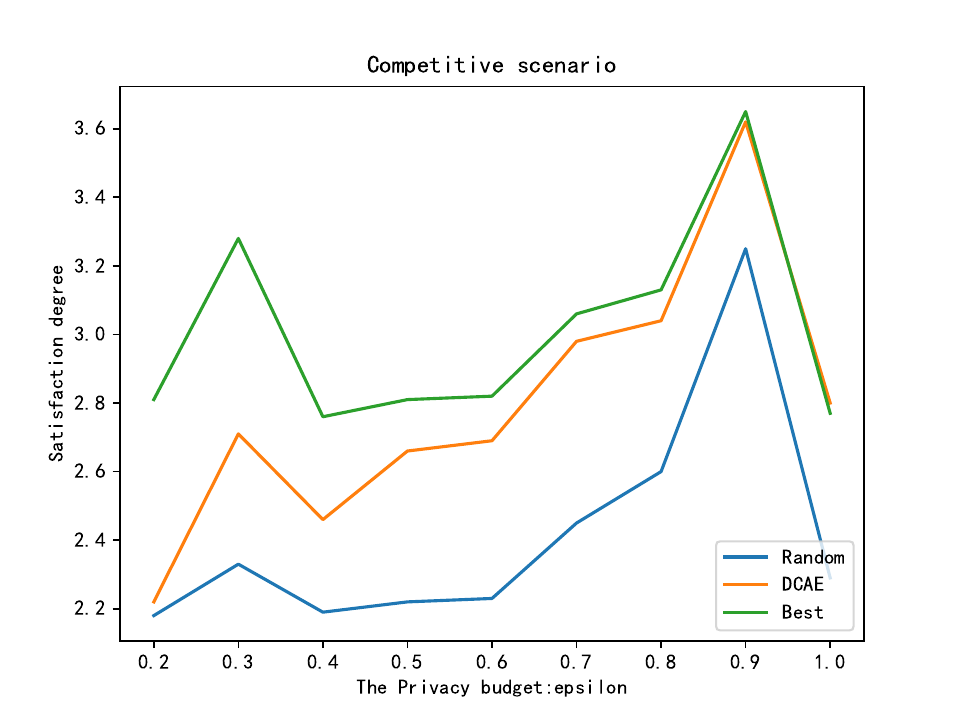}}
\caption{Changes in Mean Revenue (a) and Mean Satisfaction (b)}
\label{fig:exp_8}
\end{figure}

In a competitive environment, both DCAE and Best exhibit a gradually decreasing trend in terms of revenue due to severe restrictions on the number of datasets. However, the gap between the two is getting smaller, indicating that DCAE can still provide better privacy protection in competitive scenarios, and the decrease in revenue is related to the sharp decline in the number of datasets. In terms of user satisfaction, the Random curve is obviously the lowest. Even in the case of declining revenue, DCAE can still maintain a high level of satisfaction, and the gap with Best becomes closer as the level of privacy protection increases. It can be seen that when the privacy budget reaches its maximum, DCAE can even achieve the same level of satisfaction as Best. Therefore, it can be seen that DCAE can still maintain a good level of privacy protection even when the number of datasets is limited.
\section{Conclusion and Future Work}\label{sec:conclusion}

In this paper, we proposed a combinatorial auction mechanism called \textit{Data Trading Combination Auction Mechanism  based on the exponential mechanism} (DCAE) for data trading scenarios. To ensure privacy protection, authenticity, and high revenue, we used the exponential mechanism to select the final settlement price vector based on a specific probability distribution. We conducted simulation experiments for two different scenarios, and the results showed that compared to random selection, DCAE can achieve better revenue while ensuring the privacy of buyers. Moreover, in the non-competitive scenario, DCAE can bring even better revenue performance. In future work, several aspects are worth exploring, such as the privacy of datasets, the timeliness of data, and the honesty of buyers.


\begin{thebibliography}{10}
\expandafter\ifx\csname url\endcsname\relax
  \def\url#1{\texttt{#1}}\fi
\expandafter\ifx\csname urlprefix\endcsname\relax\def\urlprefix{URL }\fi
\expandafter\ifx\csname href\endcsname\relax
  \def\href#1#2{#2} \def\path#1{#1}\fi

\bibitem{1}
Dawex,https://www.dawex.com/en/.
\bibitem{2}
Iota,https://data.iota.org/.
\bibitem{3}
https://support.gnip.com/apis/.
\bibitem{4}
https://www.bloomberg.com/professional/product/market-data/.
\bibitem{5}
Safegraph, https://www.safegraph.com/. 
\bibitem{6}
Google bigquery, https://cloud.google.com/bigquery/.
\bibitem{7}
L. Chen, P. Koutris, and A. Kumar. Towards model-based pricing for machine
learning in a data marketplace. In SIGMOD, pages 1535–1552, 2019.
\bibitem{8}
Zhang, Y., Wang, Y., Liu, Q., Wang, X.,Jin, H. (2021). Dealer: An End-to-End Model Marketplace with Differential Privacy. Proceedings of the 27th ACM SIGKDD Conference on Knowledge Discovery and Data Mining, 3529-3538. doi: 10.1145/3447548.3467330.
\bibitem{9}
C. Dwork, “Differential privacy ,” in International Colloquium on
Automata, Languages, and Programming, 2006, pp. 1–12.
\bibitem{10}
F. McSherry and K. Talwar, “Mechanism design via differential
privacy ,” in Foundations of Computer Science, 2007. FOCS’07. 48th
Annual IEEE Symposium on. IEEE, 2007, pp. 94–103.

\bibitem{11}
R. Zhu, Z. Li, F. Wu, K. Shin, and G. Chen, “Differentially private
spectrum auction with approximate revenue maximization,” in
Proceedings of the 15th ACM international symposium on mobile ad
hoc networking and computing. ACM, 2014, pp. 185–194.

\bibitem{12}
H. Jin, S. Lu, B. Ding, K. Nahrstedt, and N. Borisov , “Enabling
privacy-preserving incentives for mobile crowd sensing systems,”
in IEEE International Conference on Distributed Computing Systems,
2016.

\bibitem{13}
Jiang X, Sun Y, Liu B, et al. Combinatorial double auction for resource allocation with differential privacy in edge computing[J]. Computer Communications, 2022, 185: 13-22.
\bibitem{14}
Wang Q, Ren K, Meng X. When cloud meets ebay: Towards effective pricing for cloud computing[C]//2012 Proceedings IEEE INFOCOM. IEEE, 2012: 936-944.
\bibitem{15}
Zaman S, Grosu D. Combinatorial auction-based allocation of virtual machine instances in clouds[J]. Journal of Parallel and Distributed Computing, 2013, 73(4): 495-508.
\bibitem{16}
Qin T, He F, Shi D, et al. Benefits of permutation-equivariance in auction mechanisms[J]. arXiv preprint arXiv:2210.05579, 2022.
\bibitem{17}
Zheng X. Data trading with differential privacy in data market[C]//Proceedings of 2020 the 6th International Conference on Computing and Data Engineering. 2020: 112-115..
\bibitem{18}
Zheng S, Cao Y, Yoshikawa M. Trading data with personalized differential privacy and partial arbitrage freeness[J]. arXiv Preprint, arXiv: 2105.01651, 2021.
\bibitem{19}
Niu C, Zheng Z, Wu F, et al. Trading data in good faith: Integrating truthfulness and privacy preservation in data markets[C]//2017 IEEE 33rd International Conference on Data Engineering (ICDE). IEEE, 2017: 223-226.
\bibitem{20}
Chen Z, Chen L, Huang L, et al. On privacy-preserving cloud auction[C]//2016 IEEE 35th Symposium on Reliable Distributed Systems (SRDS). IEEE, 2016: 279-288.
\bibitem{21}
Cheng K, Shen Y, Zhang Y, et al. Towards efficient privacy-preserving auction mechanism for two-sided cloud markets[C]//ICC 2019-2019 IEEE International Conference on Communications (ICC). IEEE, 2019: 1-6.
\bibitem{22}
Ni T, Chen Z, Chen L, et al. Differentially private combinatorial cloud auction[J]. IEEE Transactions on Cloud Computing, 2021.
\bibitem{23}
Chen Z, Ni T, Zhong H, et al. Differentially private double spectrum auction with approximate social welfare maximization[J]. IEEE Transactions on Information Forensics and Security, 2019, 14(11): 2805-2818.
\bibitem{24}
Gao G, Xiao M, Wu J, et al. DPDT: A differentially private crowd-sensed data trading mechanism[J]. IEEE Internet of Things journal, 2019, 7(1): 751-762.
\bibitem{25}
Jin X, Zhang Y. Privacy-preserving crowdsourced spectrum sensing[J]. IEEE/ACM Transactions on Networking, 2018, 26(3): 1236-1249.
\bibitem{26}
Guo J, Ding X, Wang T, et al. Combinatorial resources auction in decentralized edge-thing systems using blockchain and differential privacy[J]. Information Sciences, 2022, 607: 211-229.
\bibitem{27}
Xu Y, Xiao M, Liu A, et al. Edge resource prediction and auction for distributed spatial crowdsourcing with differential privacy[J]. IEEE Internet of Things Journal, 2022, 9(17): 15554-15569.
\bibitem{28}
Algorithmic game theory[M]. Cambridge university press, 2007.
\bibitem{29}
Gupta A, Ligett K, McSherry F, et al. Differentially private combinatorial optimization[C]//Proceedings of the twenty-first annual ACM-SIAM symposium on Discrete Algorithms. Society for Industrial and Applied Mathematics, 2010: 1106-1125.
\bibitem{30}
Dwork C. Differential privacy: A survey of results[C]//Theory and Applications of Models of Computation: 5th International Conference, TAMC 2008, Xi’an, China, April 25-29, 2008. Proceedings 5. Springer Berlin Heidelberg, 2008: 1-19.
\end{thebibliography}

\end{document}